\documentclass{cimento}
\usepackage{graphicx}% Include figure files
\usepackage{dcolumn}% Align table columns on decimal point
\usepackage{bm}% bold math
\usepackage{epsfig}
\usepackage{xspace}
\usepackage{multirow}
\usepackage{amssymb}   % for math
\usepackage{psfrag}

\renewcommand{\thetable}{\arabic{table}}

\newcommand{\sigmatt}   {\mbox{${\sigma}_{t\bar{t}}$}}
\newcommand{\ttb}       {\mbox{$t\bar{t}$}\xspace}
\newcommand{\ppbar}     {\mbox{$p\bar{p}$}\xspace}
\newcommand{\pythia}    {\textsc{pythia}\xspace}
\newcommand{\alpgen}    {\textsc{alpgen}\xspace}
\newcommand{\singletop}   {\sc{singletop}}

\newcommand{\met}       {\not{\!\!\!E_T}}
\newcommand{\ljets}     {\mbox{$\ell$+jets}\xspace}
\newcommand{\ejets}     {\mbox{$e$+jets}\xspace}
\newcommand{\mujets}    {\mbox{$\mu$+jets}\xspace}

\newcommand{\rb}        {$R={\cal{B}}(t \to Wb)/{\cal{B}}(t \to Wq)$\xspace}
\newcommand{\tld}     {$\mathcal{D}$ \xspace}
\newcommand{\ct}      {$\cos{\theta}^*$}

\title{Measurement of top quark properties at D0}

\author{E.~Shabalina \from {ins:x} \\
for D0 collaboration } 
\instlist{\inst{ins:x} University of Illinois at Chicago, 
Chicago, Illinois 60607, USA }

\PACSes{\PACSit {13.85.Rm}{Limits on production of particles}
        \PACSit {13.85.Qk}{Inclusive production with identified leptons, 
	photons, or other nonhadronic particles}
        \PACSit {14.65.Ha}{Top quarks}
        \PACSit {14.70.Ha}{W bosons}
}

\psfrag {LD2} {$\boldmath \cal D$}
\begin{document}

\maketitle

\begin{abstract}
We present a summary of the measurements of the top quark 
production and decay properties performed by the D0 experiment 
at the Fermilab 
Tevatron proton-antiproton collider at $\sqrt{s}=1.96$~TeV 
using 1 ${\rm{fb}}^{-1}$ of data. We discuss the 
first simultaneous measurement of the ratio of branching 
fractions, $R={\cal{B}}(t \rightarrow Wb)/{\cal{B}}(t \rightarrow Wq)$, 
with $q$ being a $d$, $s$, or $b$ quark, and the top quark pair production 
cross section $\sigma_{t\bar{t}}$, the first model-independent measurement 
of the helicity of $W$ bosons produced in the top quark decays, and the first 
measurement of the integrated forward-backward charge asymmetry in the 
production of \ttb pairs.   

\end{abstract}

\section{Introduction} 
The large sample of top-antitop quark pairs produced by the Fermilab 
Tevatron collider allows 
not only to perform precision measurements of such fundamental top quark 
characteristics as its production cross section and mass, but also to 
study a broad variety of top quark production and decay properties 
to address the question whether the top quark is indeed the particle predicted
by the standard model (SM). Any deviation from the SM prediction would be 
an indication of the new physics. In this paper, we study the properties 
of top quark decay by measuring the ratio of top quark branching 
fractions, \rb \cite{rb}, and the W boson helicity \cite{wh}. We 
also report on a measurement of forward-backward ($fb$) charge asymmetry in 
\ttb production \cite{asym}. 

\section{\boldmath Selection of ${t\bar{t}}$ candidates}

To study top quark properties we mainly use the top quark pair 
decay channel where one $W$ boson from a top quark  
decays into two quarks, and the other  
one into an electron or muon and a neutrino,   
referred to as the lepton plus jets (\ljets) channel.  
We select a data sample enriched in $t\bar{t}$ events by  
requiring a lepton and a jet at the trigger level. We further require 
at least four jets with transverse momentum $p_T>20$~GeV 
and pseudorapidity $|\eta|<2.5$, 
one isolated electron (muon) with $p_T>20$~GeV and 
$|\eta|<1.1$ ($|\eta|<2.0$), no other 
isolated lepton with $p_T>15$~GeV, and missing transverse energy 
$\met>20$~GeV (\ejets) ~or $\met>25$~GeV (\mujets). 
The leading jet $p_T$ is required to exceed  
$40$~GeV. 

The dominant background in the selected sample is the production of $W$ 
bosons in association with heavy and light flavor jets ($W$+jets). 
Smaller contributions arise 
from $Z$+jets, diboson and single top quark production. 
We model $W$/$Z$+jets production with the  
{\alpgen}~\cite{alpgen} Monte Carlo (MC) generator for the matrix element 
calculation and {\pythia}~\cite{pythia} for parton showering and 
hadronization. We use {\pythia} and {\singletop}~\cite{singletop} 
event generators to model diboson processes and single top quark production,
respectively. The sample also includes 
contribution from multijet events in which 
a jet is misidentified as an electron ($\ejets$ channel) or in which
a muon originating from either a semileptonic decay of a heavy quark
or an in-flight pion or kaon decay in a light flavor jet 
appears isolated ($\mujets$ channel). We determine multijet contribution 
from data. Details of the object identification, 
selection and background evaluation in \ljets channel can be found elsewhere
\cite{ljets}.  

To improve the purity of the samples used for top quark property 
measurements and/or discriminate \ttb signal from background we use 
information about the presence of $b$ jets in the 
event. We identify $b$ jets using the algorithm which 
combines variables that characterize the presence and properties of
secondary vertices and tracks with high impact parameter inside the jet 
into a neural network \cite{NNb}. 

We also utilize kinematic information to separate \ttb events 
in the selected samples from the background. We 
consider kinematic variables which discriminate between the \ttb signal 
and background and combine them into a discriminant function. 
The selected variables are required 
to be well described by the background model.     
Neglecting the correlations between the input variables, the 
discriminant function can be approximated by the expression \cite{ljets}:  
\begin{eqnarray*}
\label{eq:discr1}
{\mathcal D} &=& \frac{\prod_i s_i(x_i)/b_i(x_i)}{1+\prod_i s_i(x_i)/b_i(x_i)} \;,
\end{eqnarray*}
where $s_i(x_i)$ and $b_i(x_i)$ are the normalized distributions of variable $i$ 
for signal and background, respectively.
As constructed, the discriminant function peaks near zero for the background, and near 
unity for the signal. We use the discriminant distribution either to 
determine the fraction of \ttb events in a sample via a Poisson 
maximum-likelihood fit to data of a 
sum of signal and background discriminant
distributions \cite{rb,asym}, or to improve the purity of the samples 
by selecting events with high values of $\mathcal D$ \cite{wh}.  

\section{\boldmath Simultaneous measurement of the ratio of branching 
fractions $R={\cal{B}}(t \rightarrow Wb)/{\cal{B}}(t \rightarrow Wq)$ and 
\sigmatt.}

Within the SM top quarks decay almost exclusively to a $W$ boson and a $b$ quark.
This prediction is based on the requirements that there are three fermion families 
and the CKM matrix ~\cite{CKM} is unitary. If these assumptions are relaxed,
CKM matrix element $|V_{tb}|$ is essentially unconstrained, $|V_{tq}|$ elements can 
significantly deviate from their SM values, and 
the ratio $R$ of the top quark branching fractions, which 
can be expressed as 
\begin{eqnarray*}
\label{eq:Rdef}
R = \frac{{ \cal B}(t \rightarrow Wb)}{{ \cal B}(t \rightarrow Wq)} & = &
\frac{\mid V_{tb}\mid^2}{\mid V_{tb}\mid^2 + \mid V_{ts}\mid^2 + \mid
V_{td}\mid^2} \;,
\end{eqnarray*}
can be different from unity. A precise measurement of $R$ 
allows to perform direct measurement, free of assumptions about the
number of quark families or the unitarity of the CKM matrix,  
of the $|V_{tq}|$ elements  
via the combination with measurements of the  
single top quark production rates in $s$ and $t$ channels \cite{Vtb_talk}.  

\begin{figure}[!h!tbp]
\begin{center}
\setlength{\unitlength}{1.0cm}
\begin{picture}(18.0,5.0)
\put(0.8,0.2){\includegraphics[width=5.0cm]{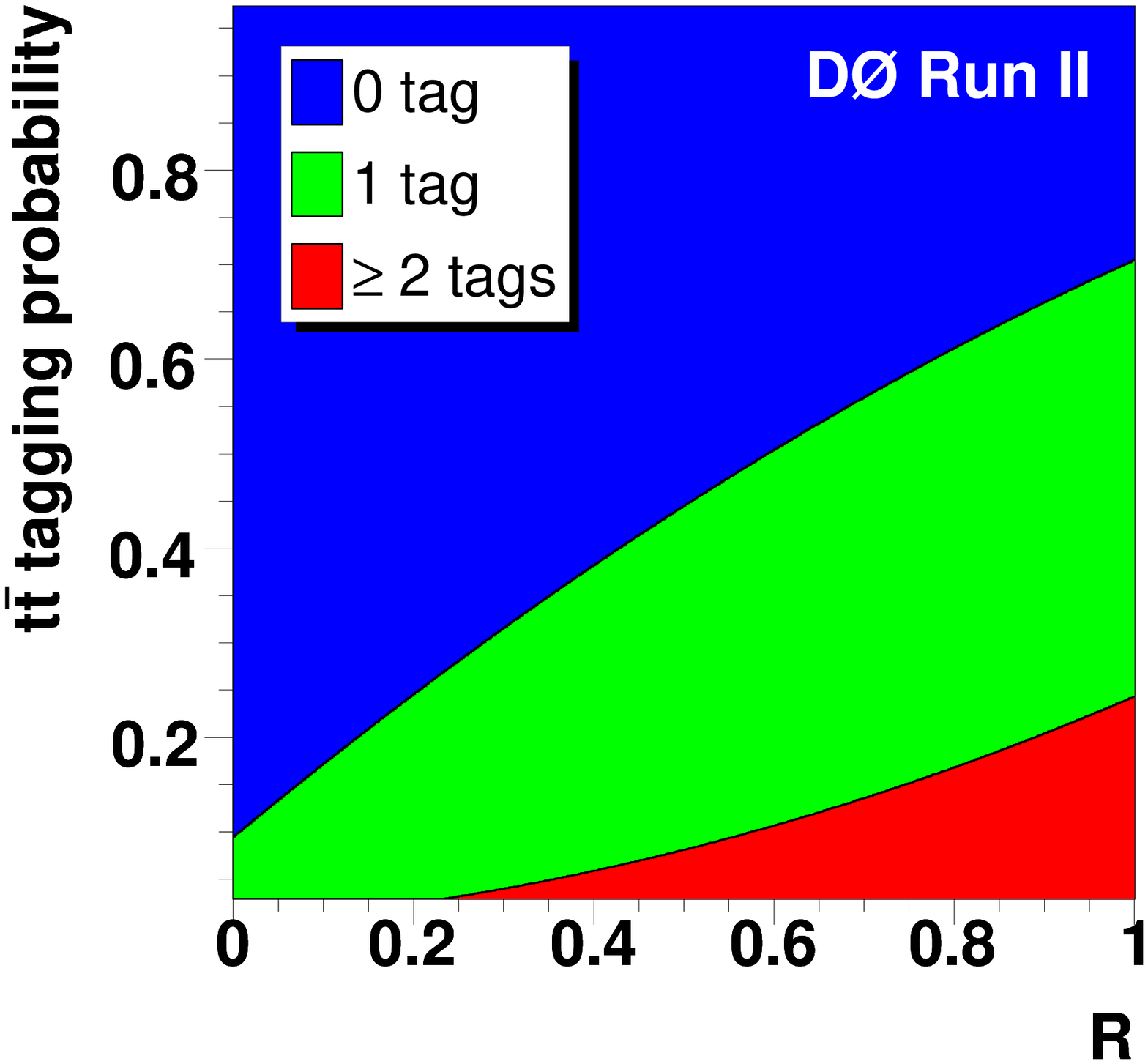}  }
\put(7.4,0.2){\includegraphics[width=5.0cm]{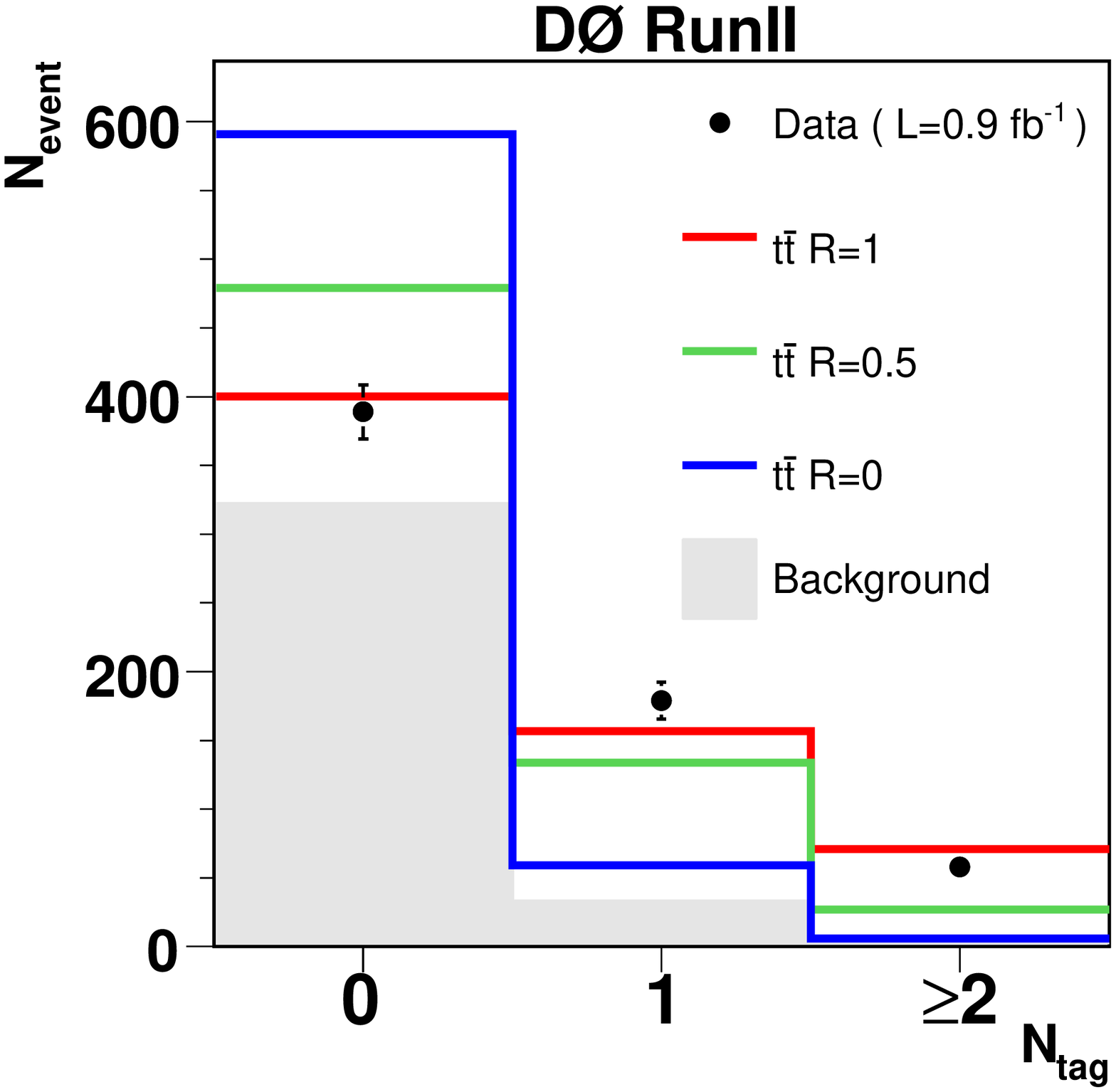} }

\put(0.2,3.8){(a)}
\put(6.8,3.8){(b)}
\end{picture}
%\vspace{-0.8cm}
\caption{\label{fig:prob} (a) Probability of 
$t\bar{t}$~events to have 0, 1 and $\ge 2$ $b$ tags as a function 
of $R$ for events with $\ge 4$ jets; (b) predicted sum of \ttb signal 
and background for $R$ = 0, 0.5 and 1 compared to data in  
the 0, 1 and $\ge 2$ $b$ tag samples for events with $\ge 4$ jets.}
\end{center}
\end{figure}

The probability for a \ttb event to have certain number of identified $b$ jets 
depends on the jet flavor, and therefore it depends on $R$.  
Fig~\ref{fig:prob}(a) shows the fraction of events with 0, 1 and 
$\ge 2$ tagged jets as a function of $R$ for \ttb events with $\ge 4$ jets.  
If $R$ is close to zero, {\it i.e.}, top quarks decay predominantly 
to $W$ bosons and light quarks, the probability to have a \ttb event with two $b$ tags is
negligible while the probability to have no tags is close to 90\%. 
Fig~\ref{fig:prob}(b) shows the comparison of the number of data events with 
0, 1 or $\ge 2$ $b$ tags to the prediction for the sum of background and 
\ttb signal with $R=0$, $R=0.5$ and $R=1$.    

%\begin{figure}[!h!tbp]
       
\begin{figure}[ht]
\begin{center}
\setlength{\unitlength}{1.0cm}
\begin{picture}(18.0,4.0)
\put(0.6,0.2){\includegraphics[width=4.2cm]{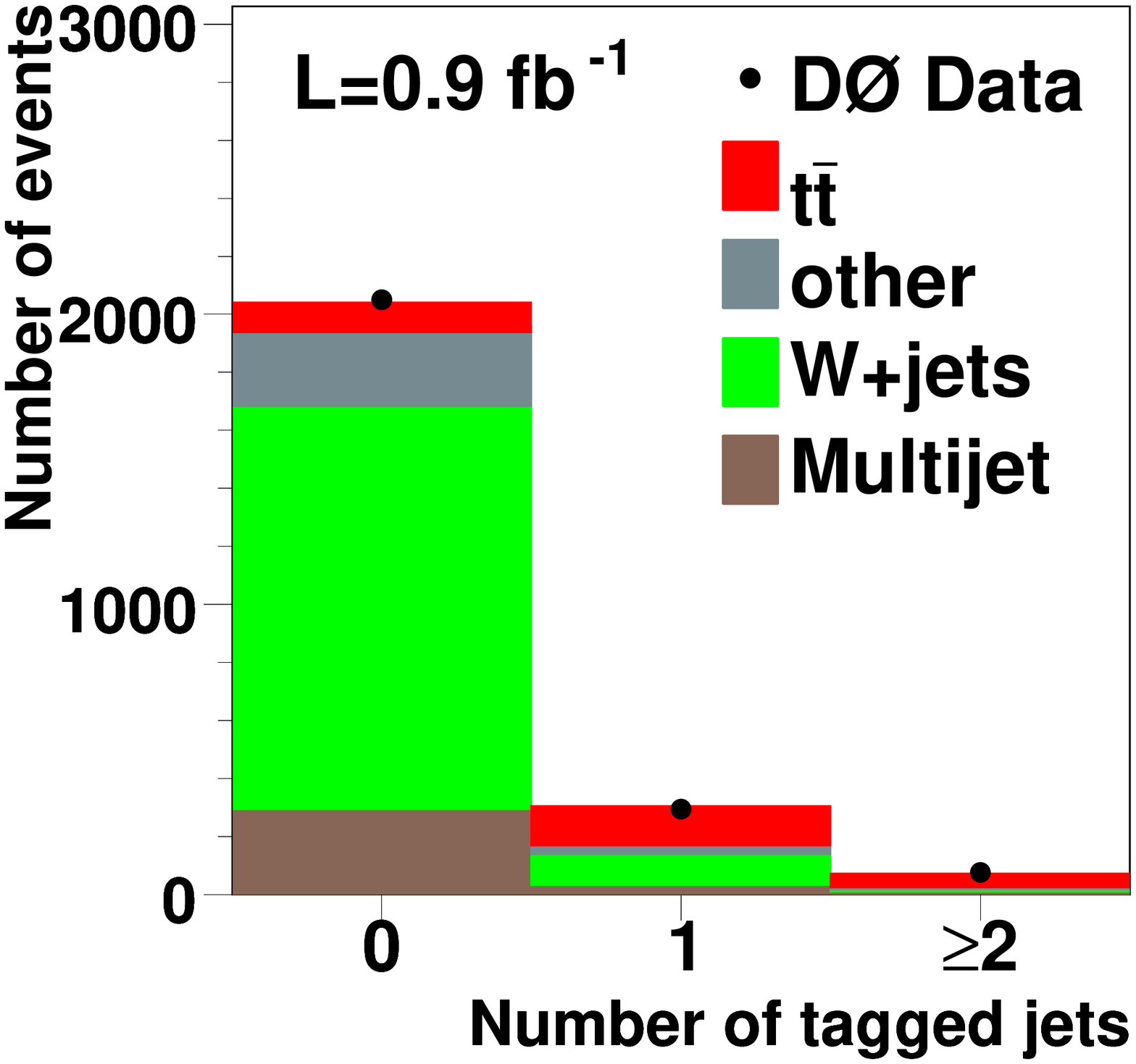}  }
\put(5.4,0.2){\includegraphics[width=4.2cm]{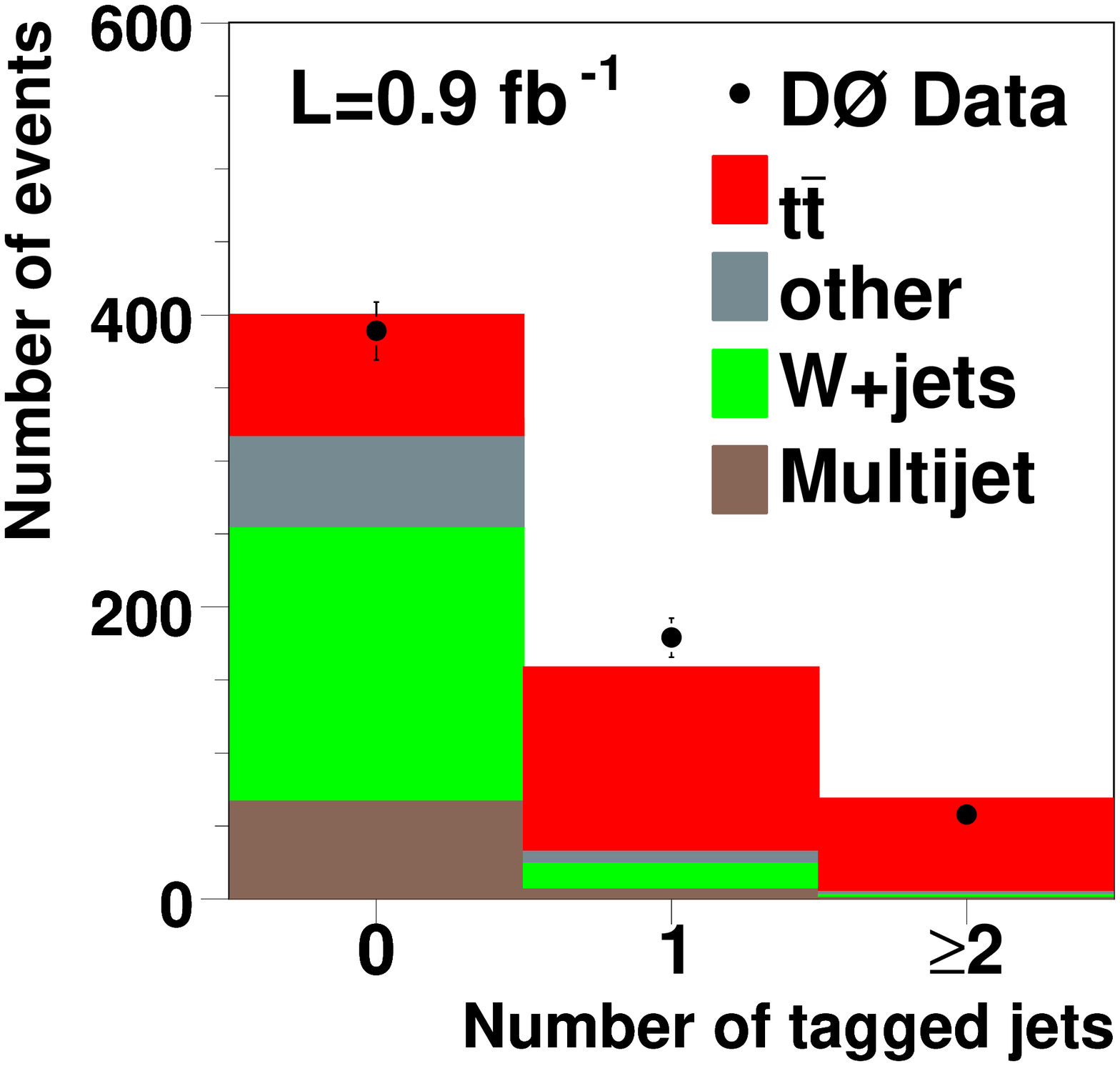} }
\put(10.0,0.2){\includegraphics[width=4.2cm]{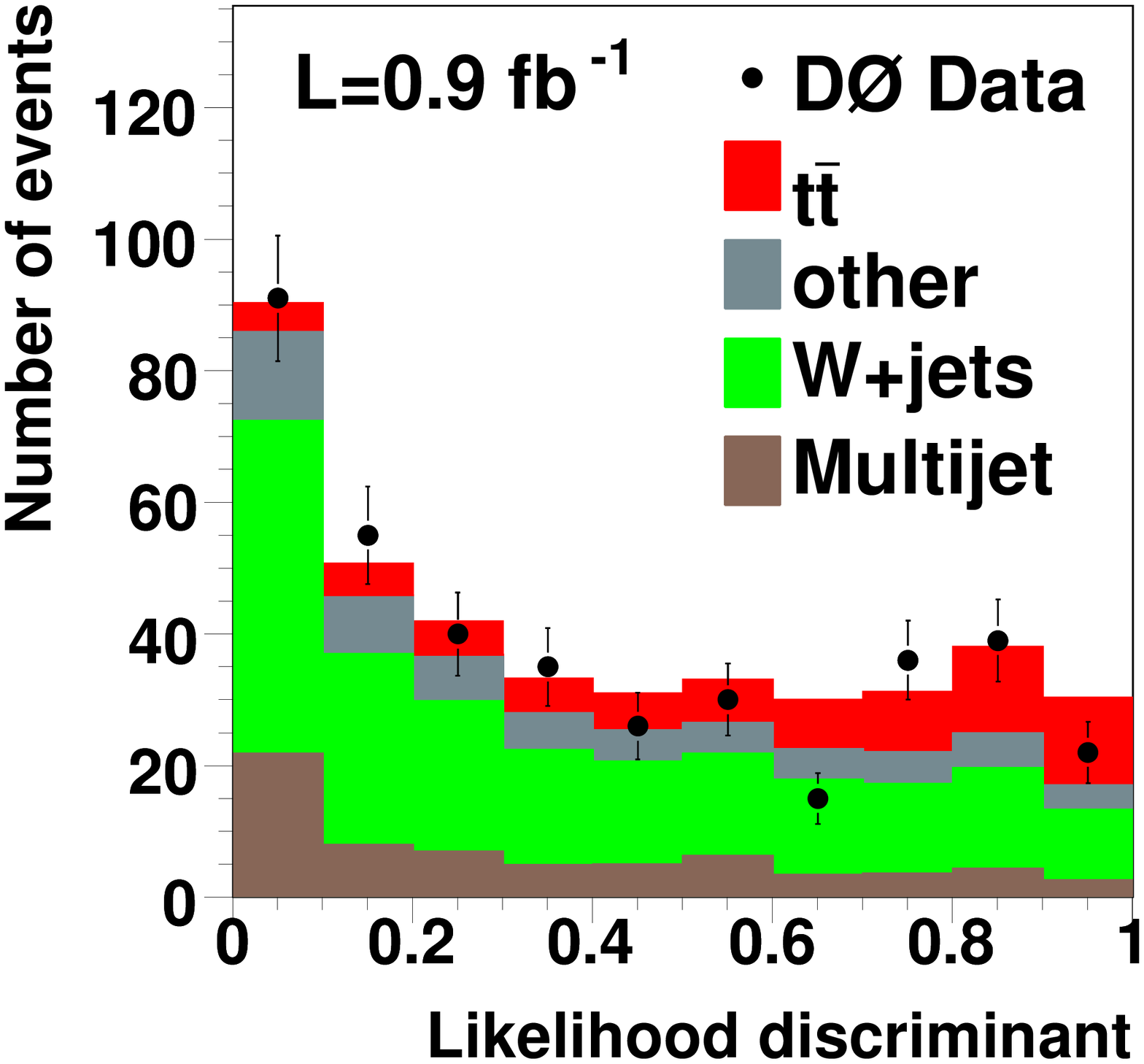} }

\put(0.0,3.8){(a)}
\put(4.8,3.8){(b)}
\put(9.5,3.8){(c)}
\end{picture}
%\vspace{-0.8cm}
\caption{\label{fig:tagbins} Predicted and observed number of events 
for the measured $R$ and $\sigma_{t\bar{t}}$ in  
the 0, 1 and $\ge 2$ $b$ tag samples for events with 
(a) three jets, (b) $\ge 4$ jets, and (c) 
predicted and observed discriminant
distribution in the 0 $b$ tag sample with $\ge 4$ jets}
\end{center}
\end{figure}

We measure the ratio $R$ using \ljets events with three or more jets. 
We split this sample into subsamples according 
to lepton flavor ($e$ or $\mu$), jet multiplicity (3 or $\ge 4$ jets) and 
number of identified $b$-jets (0, 1 or $\ge 2$), and extract $R$ and 
the \ttb production cross section $\sigma_{t\bar{t}}$
simultaneously from the fit of the predicted numbers of 
\ttb signal and background events to the observed ones with  
0, 1 and $\ge 2$ $b$ tags. For the events with 
$\ge 4$ jets and 0 $b$ tags, we include the shape of a  
topological discriminant in the fit. We build the discriminant 
function using simulated $W$+jets and \ttb events.  
For the latter we obtain a distribution for each of the three decay modes of \ttb: 
$t\bar{t} \to W^{+}bW^{-}\bar{b}$,  
$t\bar{t} \to W^{+}bW^{-}\bar{q}_l$ (or $t\bar{t}\to W^{+}q_lW^{-}\bar{b}$) and 
$t\bar{t} \to W^{+}q_lW^{-}\bar{q}_l$, 
where $q_l$ denotes a light down-type  ($d$ or $s$) quark.  
The systematic uncertainties are incorporated in the fit using nuisance 
parameters~\cite{p14btag}, each 
represented by a Gaussian term in the likelihood fit. 
The result of the fit is:  
\begin{eqnarray*}
R &=&0.97^{+0.09 }_{-0.08}~\rm{ (stat+syst)~ and} \\
\sigma_{t\bar{t}} &=& 8.18^{+0.90}_{-0.84}~\rm{ (stat+syst)~} \pm 0.50 \;
\rm{~(lumi)~pb} \;,
\end{eqnarray*} 
for a top quark mass of $175$~GeV. 
Fig~\ref{fig:tagbins} compares the distribution of 
the data to the sum of predicted background and measured signal for events with
exactly three and $\ge 4$ jets. 

We extract a limit on $R$ and $|V_{tb}|$  
following the Feldman-Cousins procedure~\cite{fc_limit} 
which yields $R > 0.79$ at 95\%~C.L. and $|V_{tb}| > 0.89$ at 95\%~C.L.  
The latter limit is derived assuming a unitary CKM matrix with three 
fermion generations.   
We also determine a model-independent limit of the ratio of 
$|V_{tb}|^2$ to the off-diagonal matrix elements to be  
$\frac{\mid V_{tb}\mid^2}{\mid V_{ts}\mid^2 + \mid V_{td}\mid^2}>3.8$ at 95\%~C.L. 
   
\section{\boldmath Measurement of the $W$ helicity in top quark decays}

In the SM the $Wtb$ top quark decay vertex has a $V-A$ structure 
which defines the fractions of $W$ bosons produced in each polarization 
state to be $0.697\pm0.012$ \cite{wh1} and $1.6 \times 10^{-4}$ \cite{wh2}
for the longitudinal 
($f_0$) and right-handed ($f_+$) fractions, respectively, for a top mass 
of 172.5 GeV. Any significant deviation from these values will be a signal of new 
physics.           

\begin{figure}[hbp]
\begin{center}
\setlength{\unitlength}{1.0cm}
\begin{picture}(18.0,3.8)
\put(0.8,0.4){\includegraphics[width=5.3cm]{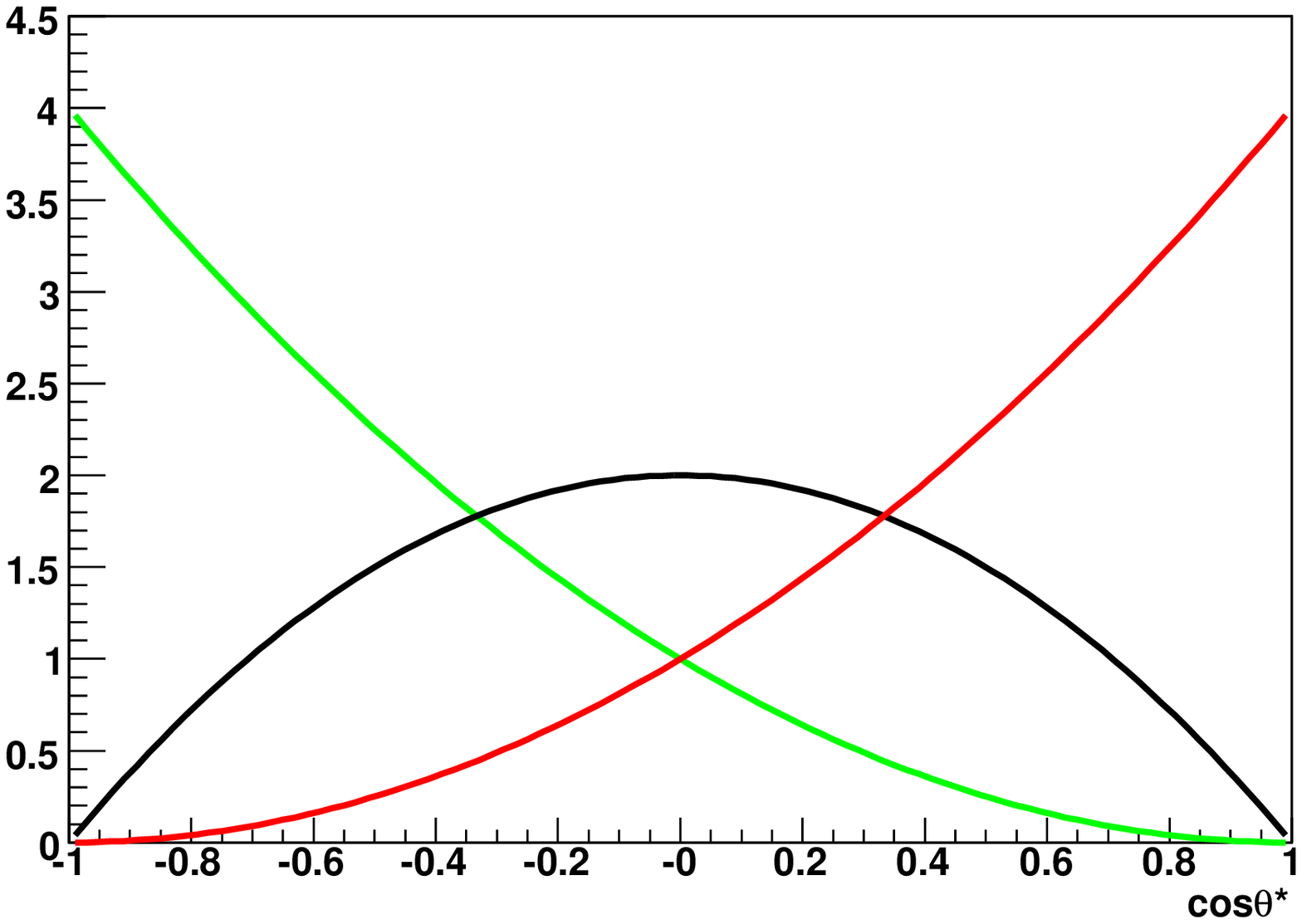}  }
\put(7.4,0.2){\includegraphics[width=5.3cm]{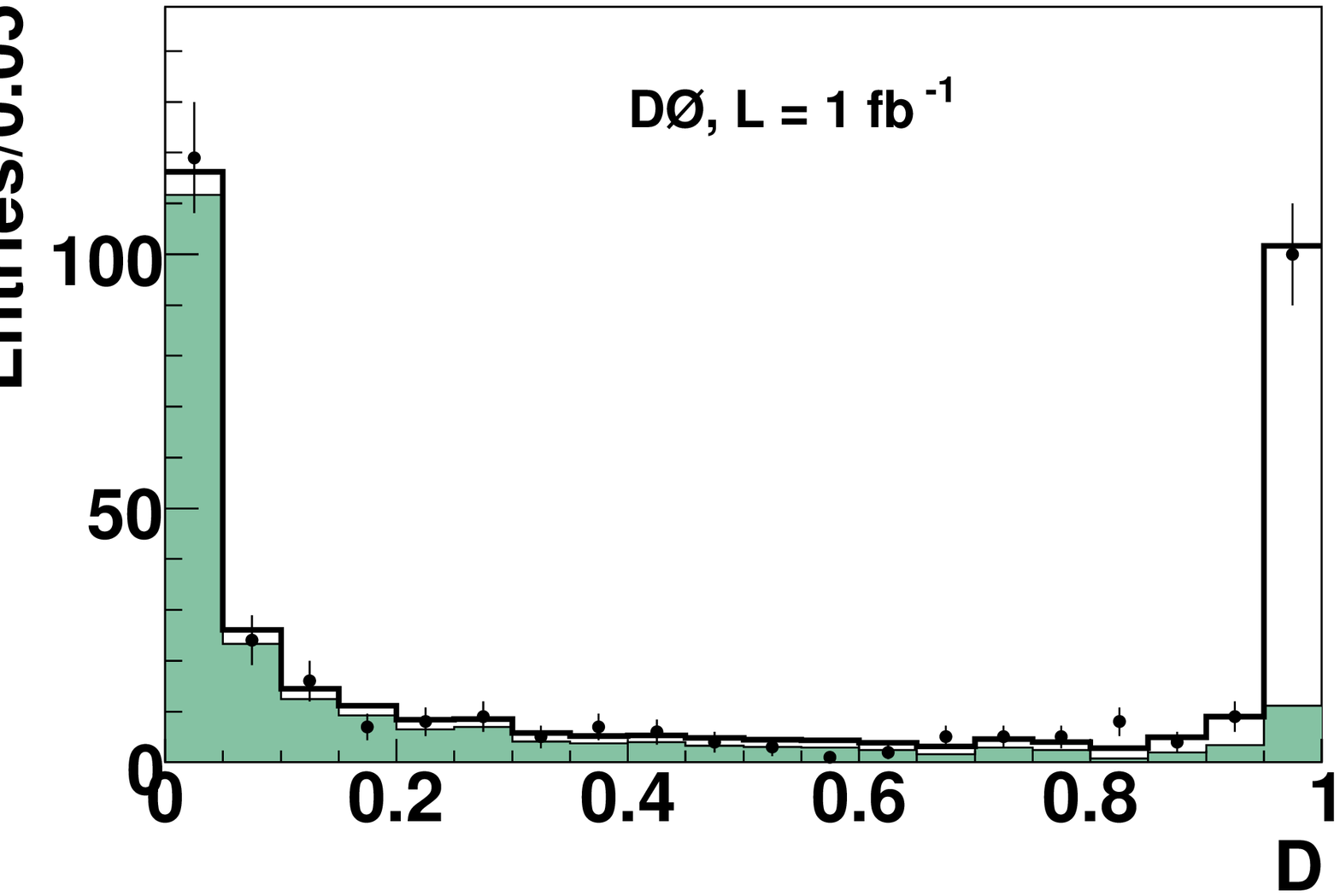} }
\put(0.2,3.4){(a)}
\put(6.4,3.4){(b)}
\end{picture}
\end{center}
\caption{(a) Distribution of the \ct ~for pure left-handed (green curve), 
right-handed (red) and longitudinal (black) $W$ boson helicity;  
(b) distribution of discriminant for signal (open histogram) and background
(green) in the e+jets channel.}    
\label{fig:wh_disc}
\end{figure}

We measure the $W$ helicity by studying the distribution of the 
angle $\theta^*$, defined in the $W$ boson rest frame,
between the momentum of down-type fermion (charged
lepton or $d$ or $s$ quark) from the $W$ boson decay and the top quark. 
The distributions of \ct ~expected for the pure   
right-handed, longitudinal and left-handed $W$'s are shown in 
fig~\ref{fig:wh_disc}(a).   

The data sample used in this analysis 
includes \ljets events and    
dilepton events where both $W$ bosons from top quarks   
decay into leptons ($e$ or $\mu$). We select such events by requiring two
charged leptons with opposite charge and $p_T>15$ GeV, and at least two jets 
with $p_T>20$ GeV. To increase signal purity we built a topological discriminant 
in each of the five channels considered ($ee$, $e\mu$, $\mu\mu$, \mujets. \ejets) 
using the kinematic variables listed in tab \ref{tab:selection}. Variables  
${\it NN}_{b_1}$ and $\langle ${\it NN}$_b \rangle$, the output value of 
the neural network of the leading jet and the mean of the 
neural network outputs of the two leading jets, respectively, include  
information about jet flavors into the discriminant. Thus, instead of 
requiring a $b$ tag in the event corresponding to a cut on ${\it NN}_b$, 
we use the full ${\it NN}_b$ distribution and avoid a loss of efficiency due to 
$b$ tag requirement. An example of the 
discriminant distribution in the \ejets channel is shown in 
fig~\ref{fig:wh_disc}(b). In each channel we select a cut on \tld to achieve 
the best expected precision for $W$ helicity. Tab \ref{tab:selection} lists 
the cut values and the composition of each sample after the cut. 

\begin{table}
\caption{\label{tab:selection}%
Summary of the multivariate selection and number of selected events for 
each of the \ttb final states in $W$ helicity measurement. The uncertainties 
are statistical only, except for the  background estimates
in the $ee$ and $\mu\mu$ channels, in which systematic uncertainties arising from imperfections in the MC model of the data are included.
}
\begin{tabular}{lccccc}
\hline 
       &  $e+$jets & $\mu+$jets & $e\mu$ & $ee$ & $\mu\mu$  \\\hline
Variables   & ${\cal C},$ $ {\cal S}$,  ${\cal A}$, $H_T$, & ${\cal C},$ $ {\cal S}$,  $H_T$,   
                    & ${\cal C},$ $ {\cal S}$,  $h$, $\met$, & ${\cal A}$,   
		    $ {\cal S}$ , $k_{T{\rm min}}^\prime$, 
		    & ${\cal A}$, $ {\cal S}$ , $h$, $m_{jj{\rm min}}$,\\
used in \tld       &   $h$,  $k_{T{\rm min}}^\prime$,     
                    &   $k_{T{\rm min}}^\prime$,     
		    &    $k_{T{\rm min}}^\prime$, {\it NN}$_{b_1}$,    
		    &    $\met$, {\it NN}$_{b_1}$,      
		    & $\chi2_Z$,    {\it NN}$_{b_1}$,   \\
	       &  $\langle ${\it NN}$_b \rangle$ &  $\langle ${\it NN}$_b \rangle$ 
	       &  $m_{\ell\ell}$ & $m_{\ell\ell}$ & $m_{\ell\ell}$ \\ [0.05 in]
	       
Cut on \tld & $>0.80$ & $>0.40$ & $>0.08$ & $>0.986$ & $>0.990$ \\
Background  &    $21.1 \pm 4.5$    &  $33.0 \pm 5.2$ & $9.9 \pm  2.5$ & $2.2 \pm 0.9$ &      $4.8 \pm 3.4$   \\
Data  &   121    & 167    &   45  & 15    & 15 \\ \hline 
\end{tabular}
\end{table}

\begin{figure}[hbp]
\begin{center}
\includegraphics[width=4.4cm]{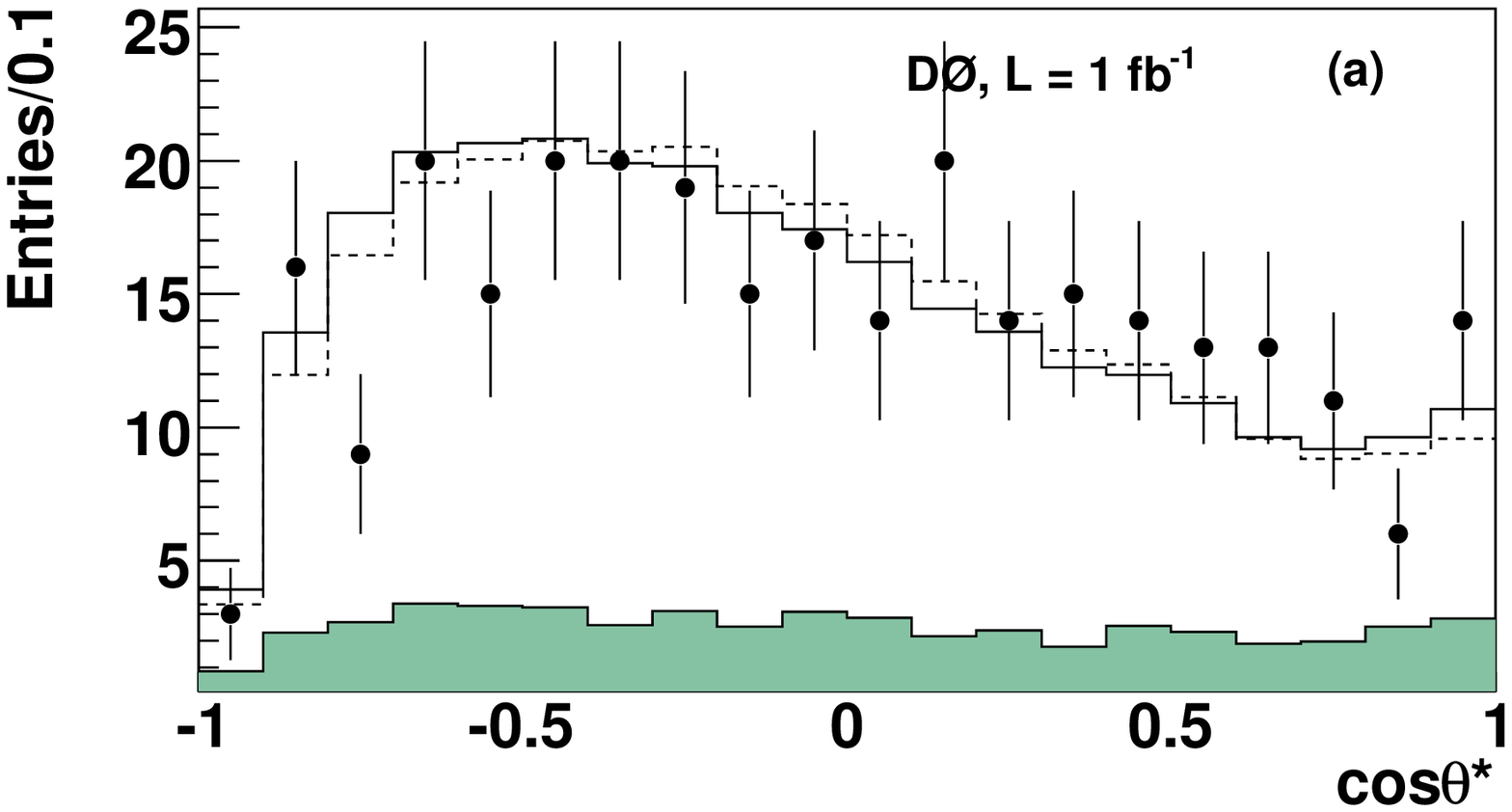}
\includegraphics[width=4.4cm]{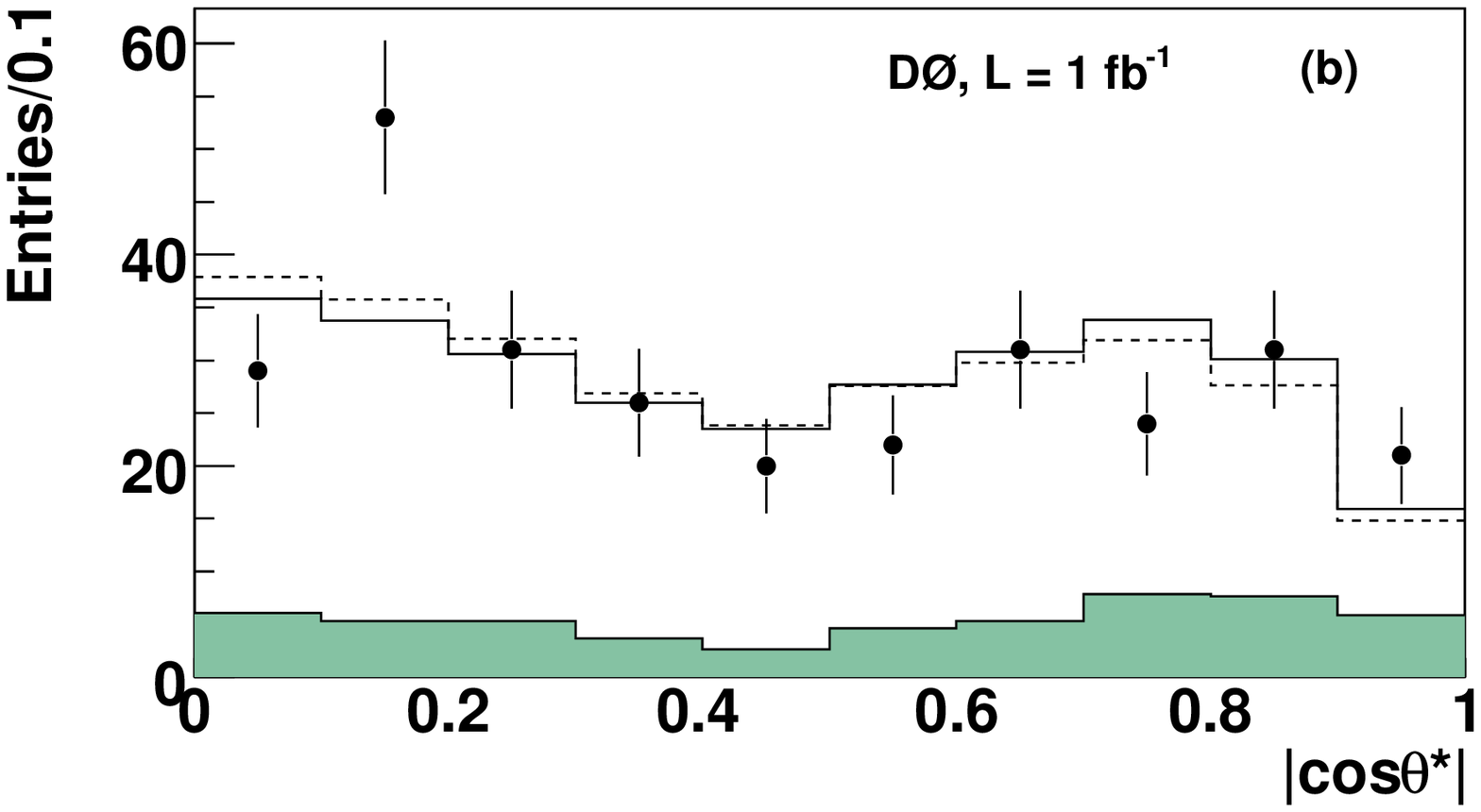}
\includegraphics[width=4.4cm]{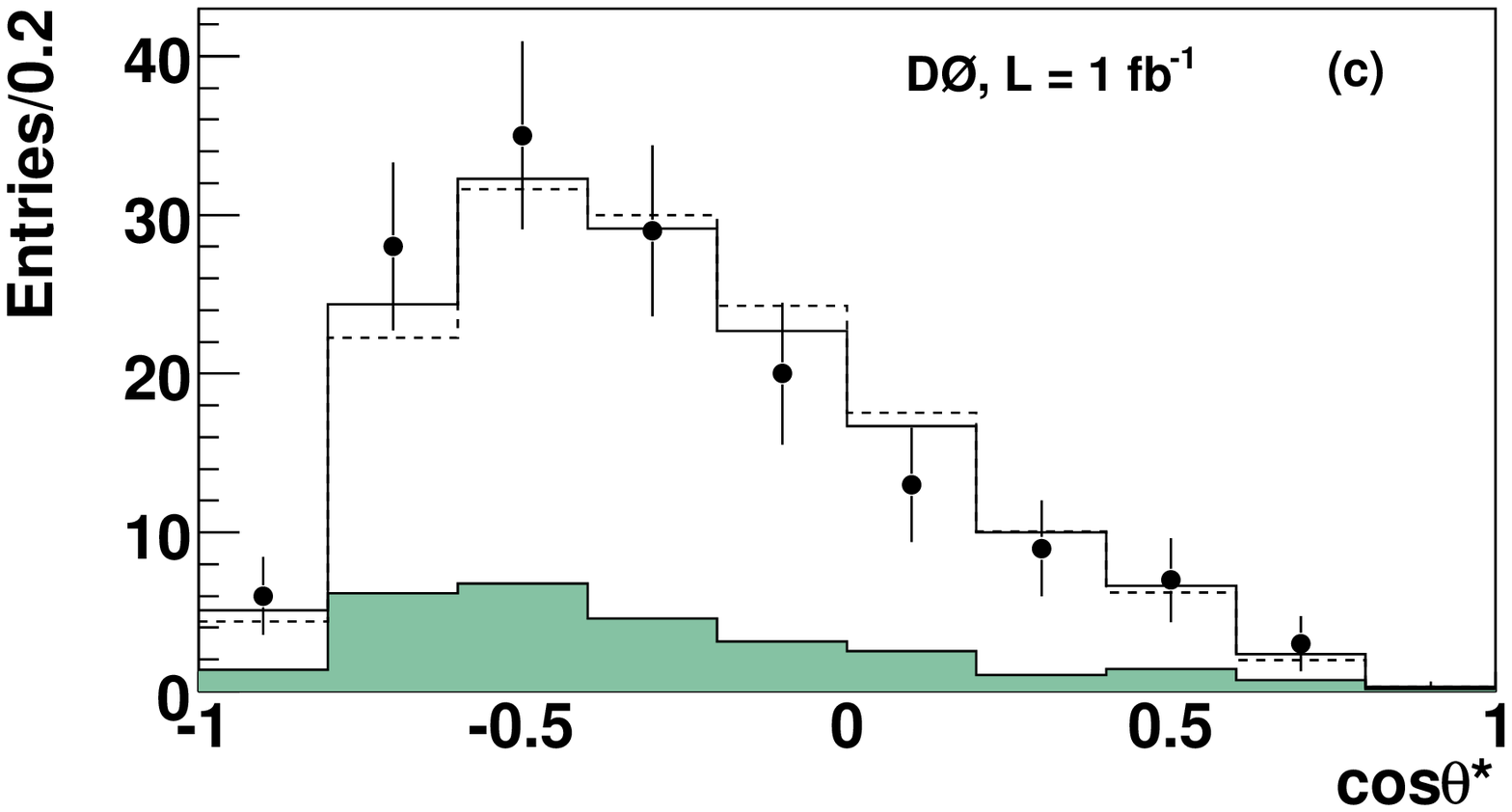}
\end{center}
\caption{Comparison of the \ct ~distributions in data (points with
error bars) with the result of the fit (solid open histograms) and the standard
model expectation (dashed open histograms) for (a) leptonic $W$ boson decays and
(b) hadronic $W$ boson decays in \ljets evens, and (c) in dilepton events.}    
\label{fig:wh_costheta}
\end{figure}

To calculate \ct ~we reconstruct events in \ljets channel using a 
kinematic fitter.
The fitter varies the four-momenta of the objects in the event within their
resolutions and minimizes a $\chi^2$ under the following constraints: 
{\it i})  both $W$ boson masses are required to be exactly 80.4 GeV;      
{\it ii}) the masses of two reconstructed top quarks are required to be 
exactly 172.5 GeV.   
Among possible 12 solutions we choose the one with the best $\chi^2$ from  
the kinematic fit and with the maximim probability based on ${\it NN}_b$ 
values of all four jets. 
In the dilepton channel, the kinematics are underconstrained. We reconstruct 
dilepton events assuming a top quark mass of 172.5 GeV. Each event
provides two entries to the \ct ~distribution. 
%We calculate \ct for
%each of the two leading jets with each of the two charged leptons associated
%with the jet resulting from four solutions. 
Fig~\ref{fig:wh_costheta} shows the reconstructed \ct ~distributions for (a) 
leptonic and (b) hadronic $W$ boson decays in \ljets channel, and (c) 
for dilepton channel. Although 
hadronic $W$ decays do not allow to discriminate between left-handed and
right-handed $W$ bosons, they do add information on the  longitudinal $W$
boson fraction. Including the hadronic decays in the measurement improved 
the sensitivity by 20\%. 

\begin{figure}
\begin{center}
\includegraphics[width=0.4\textwidth]{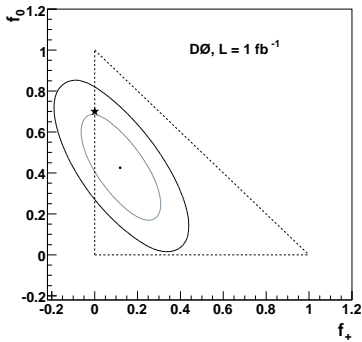}
\end{center}
\caption{Result of the model-independent $W$ boson helicity fit. The ellipses
show the 68\% and 95\% C.L. contours, the triangle corresponds to the physically
allowed region where $f_0+f_{+} \leq 1$. The star denotes the SM prediction.}
\label{fig:wh_fit}
\end{figure}

We extract $f_0$ and $f_+$ from the binned Poisson likelihood fit of the sum 
of background and   
right-handed, left-handed, and longitudinal helicity templates for signal   
to data. The measured values of $f_0$ and $f_+$ are: 
\begin{eqnarray*}
f_0 &=&0.425 \pm 0.166 ({\rm stat}) \pm 0.102 ({\rm syst}) \\
f_+ &=&0.119 \pm 0.090 ({\rm stat}) \pm 0.053 ({\rm syst})\; 
\end{eqnarray*} 
with a correlation factor of -0.83. Fig \ref{fig:wh_fit} shows the 
result of the fit along with the 68\% and 95\% C.L. contours from the fit, 
including
all uncertainties. The measured values are consistent with the SM, as there is a 30\%
chance to observe a larger discrepancy given the statistical and systematic
uncertainties of the measurement. If $f_+$ is fixed to the SM value, we find   
$f_0 =0.619 \pm 0.090 ({\rm stat}) \pm 0.052 ({\rm syst}) $  
and if $f_0$ is fixed to the SM value, $f_+$ yields
$f_+ =-0.002 \pm 0.047 ({\rm stat}) \pm 0.047 ({\rm syst})$.  

\section{\boldmath Measurement of the forward-backward charge asymmetry in \ttb production}

Leading order calculations of \ttb production in \ppbar collisions 
predict that the kinematic distributions of a top quark pair are charge 
symmetric. But this symmetry is accidental. 
At the higher orders top pair production is expected to be charge asymmetric. 
In particular, at NLO the predictions range from 5\% to 10\% \cite{NLO}. 
NNLO calculations \cite{NNLO} are not complete but deviate significantly from 
the NLO predictions. But in both cases expected asymmetry is small.  
If new physics is present in \ttb production, the 
observed asymmetry can be much larger than that predicted by the SM, thus 
making the charge asymmetry measurement a probe for new physics. 

A charge asymmetry in \ttb production can be observed as a forward-backward
integrated asymmetry defined as        
$A_{fb} = (N_f-N_b)/(N_f+N_b)$,
where $N_f(N_b)$ is the number of events with a positive
(negative) $\Delta y$ and 
$\Delta y = y_t - y_{\bar t}$ is the signed difference between top and
anti-top quark rapidities. 
The studies using events simulated with MC@NLO event 
generator \cite{MCNLO} showed that integrated asymmetry strongly depends on
the region of phase space under study. In particular, we observe large 
variations of asymmetry from +8\% to -3\% as a function of the fourth 
highest particle jet $p_T$. It also shows a strong dependence on the number 
of jets in the event. 

For this measurement we select \ljets events with at least four jets, one 
of which is required to be identified as a $b$ jet by the neural network 
algorithm. We reconstruct the kinematics of the top quark pairs using the  
kinematic fitter described above with the difference that top quark mass is 
assumed to be 170 GeV. We use the $b$-tagged and the three remaining highest
$p_T$ jets in the fit, thus reducing the number of possible parton-jet
assignments considered in the fit. We select the solution with the lowest $\chi^2$.   

We build a discriminant to separate \ttb signal from the background 
using the following variables: {\it i}) the $p_T$ of the leading $b$-tagged jet, 
{\it ii}) the $\chi^2$ of the kinematic fit, {\it iii}) the invariant mass of 
two jets assigned to $W$ boson by the fit, 
and {\it iv}) $k^{min}_T = p^{min}_T R^{min}$, where $R^{min}$ is the smallest
angular distance between any two jets used in kinematic fit, and $p^{min}_T$ is
the smaller of these two jets' transverse momenta.       

\begin{figure}[hbp]
%\psfrag{LD2} {$\bm \cal D$}
\begin{center}
\epsfig{file=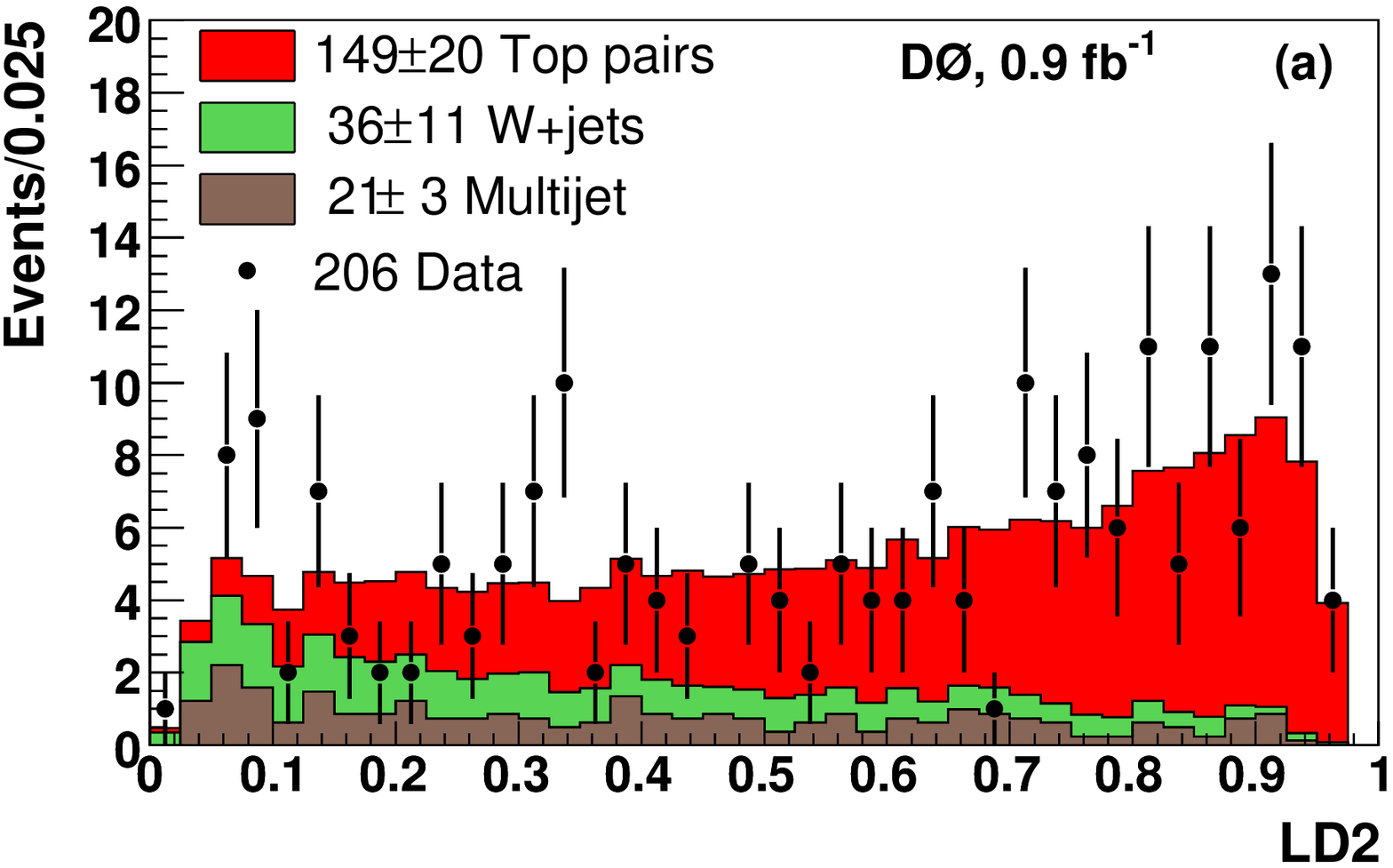,width=5.8cm}
\epsfig{file=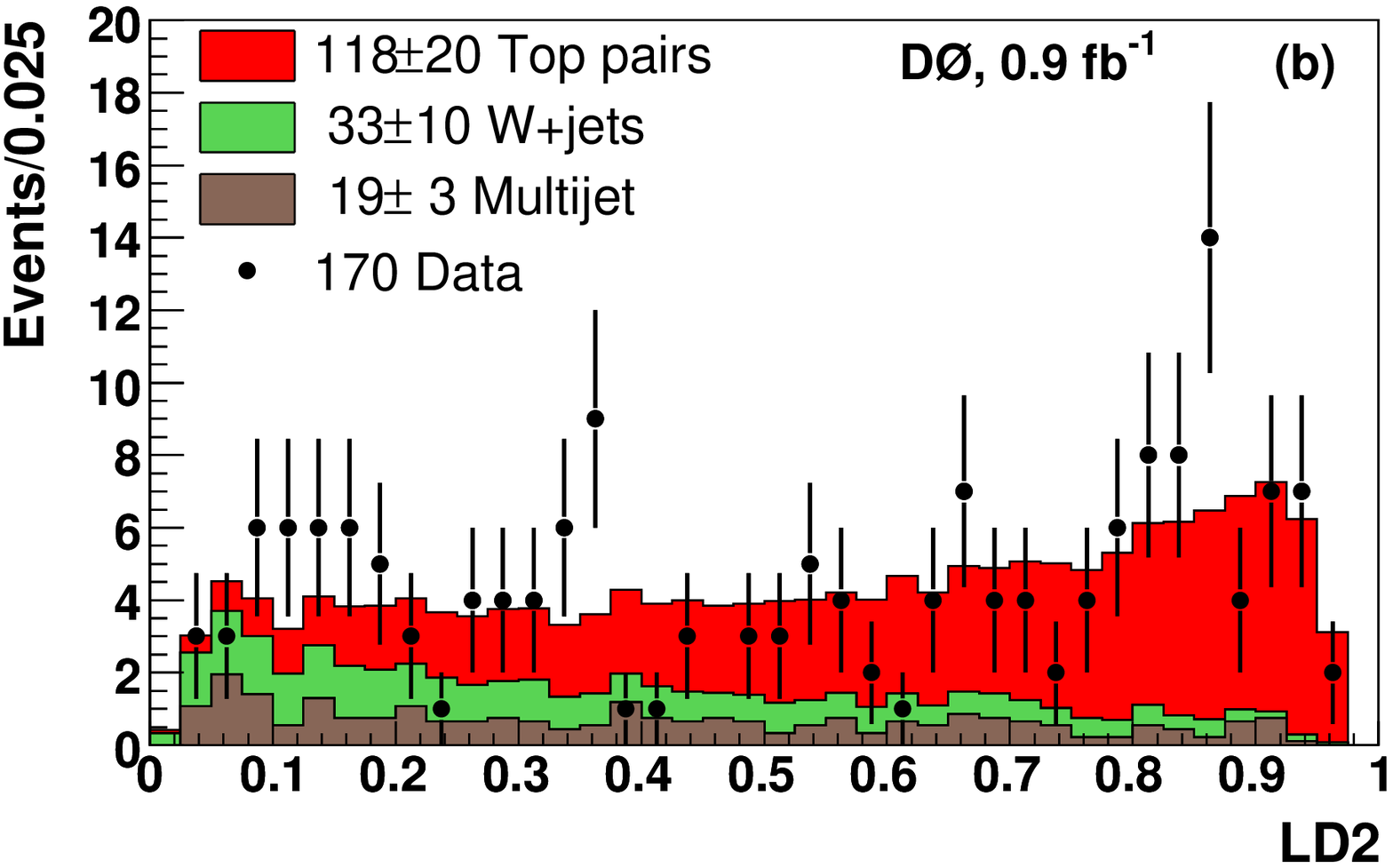,width=5.8cm}
\end{center}
\caption{Distribution of the discriminant for data and a sum of \ttb 
signal and background for events reconstructed as forward ($\Delta
y_{reco}>0$) (a) and backward ($\Delta y_{reco}<0$) (b).}    
\label{fig:asym1}
\end{figure}

We extract the \ttb content and $A_{fb}$ simultaneously by performing a 
maximum-likelihood fit of the sum of 
backgrounds and backward and forward signal to the discriminant distribution 
in data and to the sign of the reconstucted $\Delta y$. 
Forward and backward signal have the same discriminant distribution but different 
sign of the reconstructed $\Delta y$.  
The fitted sample composition compared to the discriminant distribution in data
for events with $\ge 4$ jets is shown in fig \ref{fig:asym1}. We measure
asymmetry for events with $\ge 4$, $=4$ and $\ge 5$ jets. We find
$A_{fb}=(12\pm8)$\%, $A_{fb}=(19\pm9)$\%, and $A_{fb}=(-16^{+15}_{-17})$\%,
respectively, consistent with the NLO predictions from MC@NLO. 

\begin{figure}[hbp]
\begin{center}
\setlength{\unitlength}{1.0cm}
\begin{picture}(18.0,4.3)
\put(0.4,0.3){\includegraphics[width=7.0cm]{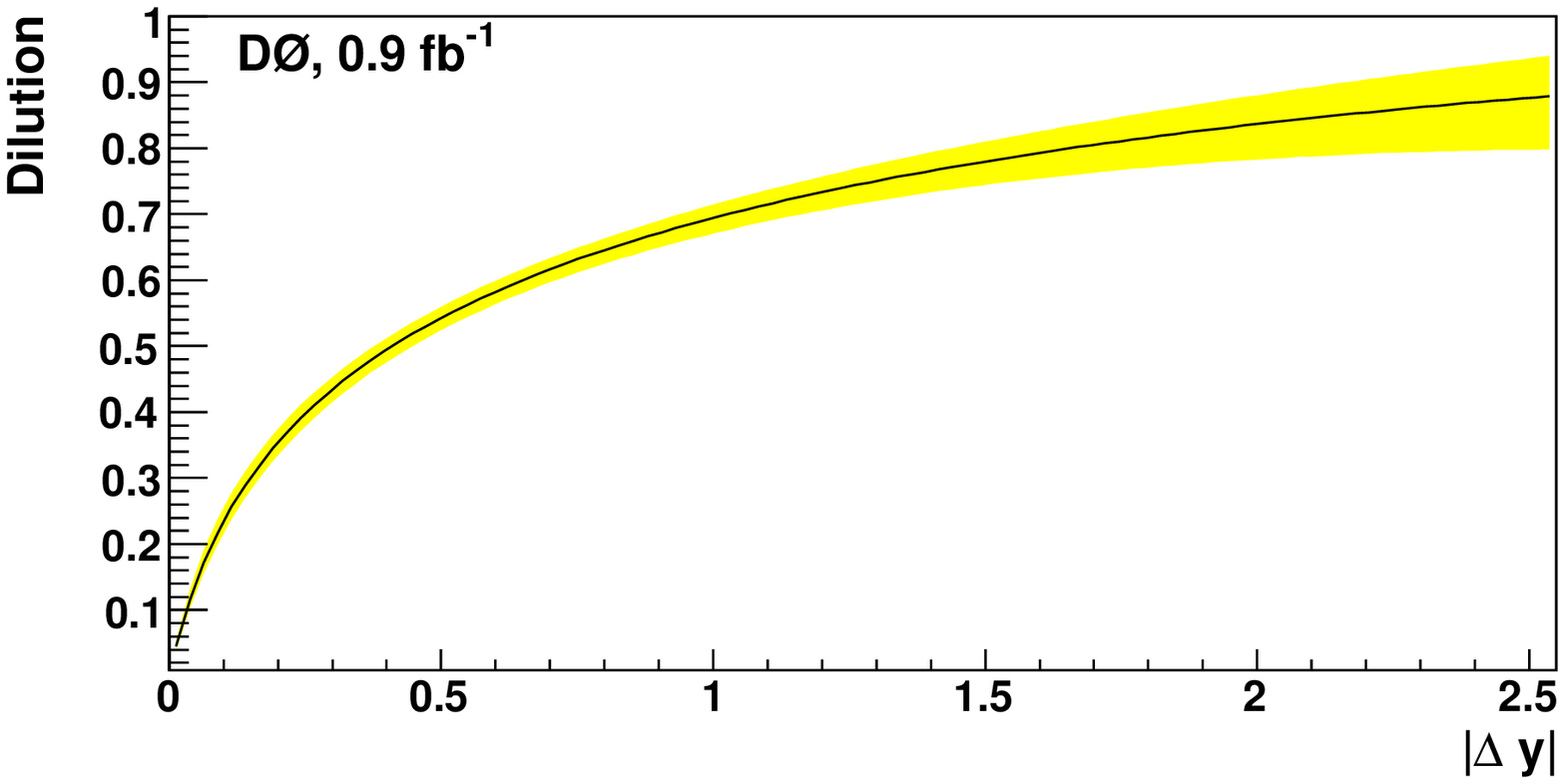}  }
\put(7.8,0.2){\includegraphics[width=5.8cm]{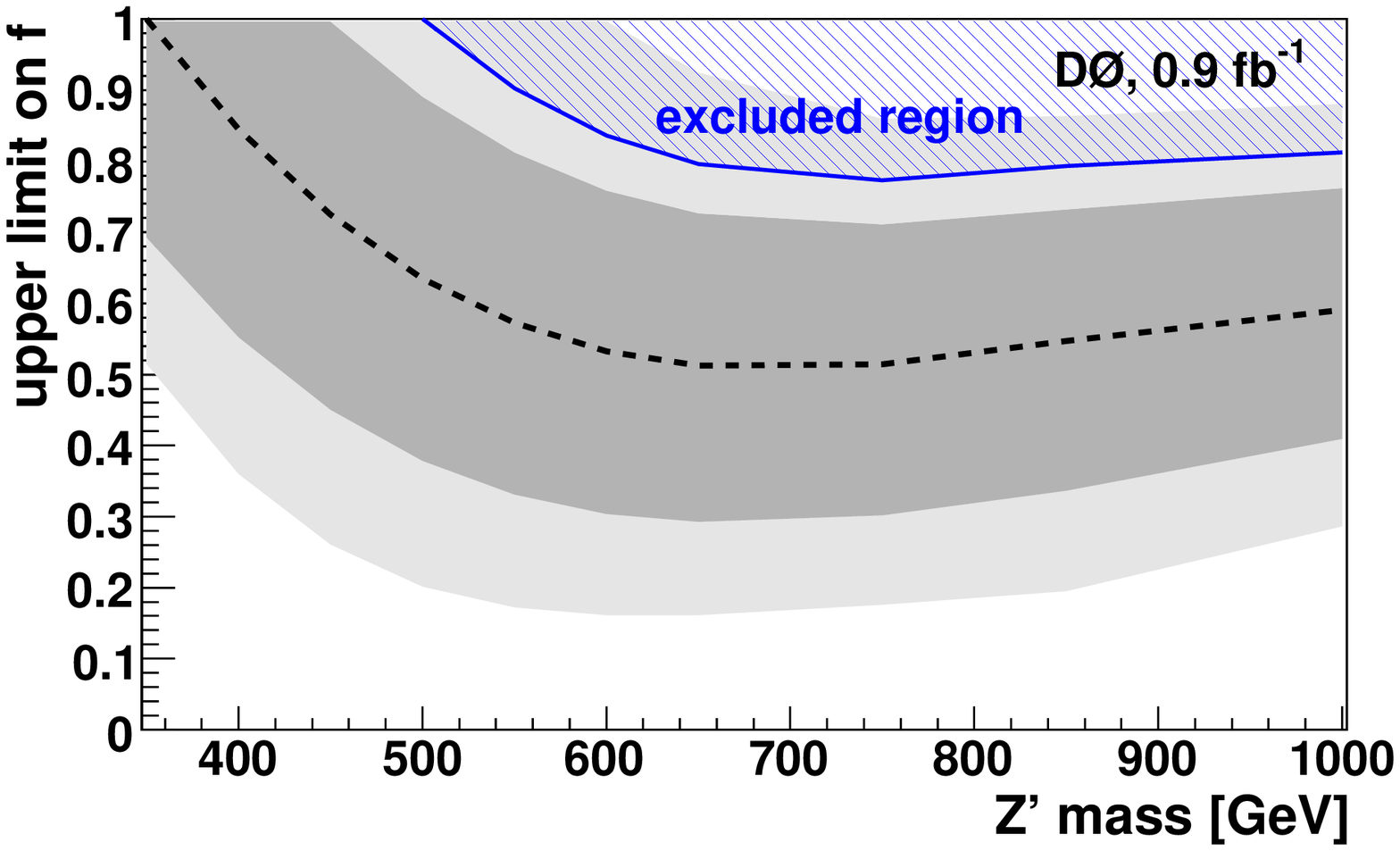} }
\put(0.2,3.8){(a)}
\put(7.4,3.8){(b)}
\end{picture}
\end{center}
\caption{(a) The geometric dilution and its uncertainty band as a function 
of generated $|\Delta y|$ for the standard model \ttb production and $\ge 4$ 
jets;
(b) 95\% C.L. limits on the fraction of \ttb produced via a $Z'$ resonance 
as a function of the $Z'$ mass. Expected (observed) limits are shown by the
dashed (solid) curve. Shaded bands correspond to one and two standard 
deviations from the expected limit. The excluded region is hatched.}     
\label{fig:asym2}
\end{figure}

To compare the measured integrated asymmetry with theory we either have to
correct it for the reconstruction and acceptance effects or provide a
prescription that would allow such comparison. We have chosen the
latter option since correction can reduce sensitivity to the effects beyond the
SM. We define the asymmetry at the particle level as             
\begin{eqnarray*}
\label{eq:asym}
A_{fb}(|\Delta y|) = \frac{g(|\Delta y|)-g(-|\Delta y|)}{g(|\Delta y|)+g(-|\Delta y|) } \; ,
\end{eqnarray*}
where $g$ is the probability density for generated $|\Delta y|$ within the 
acceptance. Since the measured asymmetry can be diluted due to misreconstruction
of the event kinematics or misidentification of the lepton charge the predicted
reconstructed asymmetry is given by     
\begin{eqnarray*}
A_{fb}^{\rm pred} = \int_{0}^{\infty} {A_{fb}(\Delta y)D(\Delta y)[g(\Delta y)+g(-\Delta y)] d
\Delta y  }
\end{eqnarray*}
where $D$ is the dilution function. It is defined as $D = 2P-1$, where $P$ is
the probability to reconstruct the correct sign of $\Delta y$. This function is
determined using simulated \ttb events passed through the full reconstruction chain. 
Dilution represents the fraction of asymmetry visible in the
detector and is parameterized as  
\begin{eqnarray*}
D(|\Delta y|) = c_0 ln(1+c_1 |\Delta y|+c_2 |\Delta y|^2) \; , 
\end{eqnarray*}
where the parameters depend on the number of jets in the event. The dependence of the
dilution function on $|\Delta y|$ introduces model dependence into any correction 
of $A_{fb}$ from the
observed to the particle-level asymmetry as they depend on the model's 
$A_{fb}(|\Delta y|)$ shape. An example of the dilution function is
shown in fig~\ref{fig:asym2}(a) and the parameters can be found in \cite{asym}. 

A contribution from the new physics can affect the \ttb forward-backward charge
asymmetry. Here we consider the sensitivity of the measurement to 
\ttb production via heavy neutral gauge bosons ($Z'$) decaying predominantly
to quarks. We study the scenario where the coupling between the $Z'$ boson and
the quarks is proportional to that of $Z$ boson and the quarks, and the
interference effects with the SM \ttb production are negligible. Unlike the
direct searches of $Z'$ \cite{reson} the asymmetry measurement is sensitive to
the production of both narrow and wide resonances. We simulated \ttb events 
produced via narrow $Z'$ resonance as in \cite{reson} using \pythia and found large positive
asymmetries ranging from 13\% to 35\% depending on the resonance mass. We
predict the distribution of $A_{fb}$ as a function of the fraction of \ttb
events produced via $Z'$ for each $Z'$ mass and set limits following the
Feldman-Cousins procedure \cite{fc_limit} (see fig \ref{fig:asym2}(b)).  
       
\section{Summary}
In summary, we have presented the measurements of the production and 
decay properties of the top quark. All measurements are consistent with the 
SM. The large data set collected at the Tevatron will allow more precise measurements 
of the top quark properties in the near future and facilitate searches for 
new physics in the top quark sector.

\acknowledgments
% acknowledgement_paragraph_r2.tex                         5/23/08
%
We thank the staffs at Fermilab and collaborating institutions, 
and acknowledge support from the 
DOE and NSF (USA);
CEA and CNRS/IN2P3 (France);
FASI, Rosatom and RFBR (Russia);
CNPq, FAPERJ, FAPESP and FUNDUNESP (Brazil);
DAE and DST (India);
Colciencias (Colombia);
CONACyT (Mexico);
KRF and KOSEF (Korea);
CONICET and UBACyT (Argentina);
FOM (The Netherlands);
STFC (United Kingdom);
MSMT and GACR (Czech Republic);
CRC Program, CFI, NSERC and WestGrid Project (Canada);
BMBF and DFG (Germany);
SFI (Ireland);
The Swedish Research Council (Sweden);
CAS and CNSF (China);
and the
Alexander von Humboldt Foundation (Germany).
%
   % input acknowledgement

\end{document}